\documentclass{dcds-B}
\usepackage{graphics}
\usepackage{epsfig}

\def\currentvolume{?}
\def\currentissue{}
\def\currentyear{2003}
\def\currentmonth{Month}
\def\ppages{??--??}

\title[Discrete models of force chain networks]
      {Discrete models of force chain networks}
\author[J.E.S. Socolar]{}

\subjclass{????}

 \email{socolar@phy.duke.edu}

\newcommand{\dy}{\raisebox{1.6ex}{\rotatebox{180}{\textsf{Y}}}}
\newcommand{\uy}{\textsf{Y}}
\newcommand{\nh}{\hat{n}}
\newcommand{\uh}{\hat{u}}

\newcommand{\ev}{\vec{e}}
\newcommand{\fv}{\vec{f}}
\newcommand{\yv}{\vec{y}}

\newcommand{\rv}{\vec{r}}

\newcommand{\tens}{\vec}

\newcommand{\Am}{\tens{M}}
\newcommand{\eqname}{Boltzmann }
\newcommand{\Eqname}{Boltzmann }

\begin{document}

\maketitle

\centerline{\scshape Joshua E.S. Socolar}
 \medskip

{\footnotesize \centerline{ Physics Department}  \centerline{ Duke University } 
  \centerline{ Durham, NC, 27708 }
}
\medskip 

\centerline{(Communicated by ???)}

\medskip

\begin{abstract}
  A fundamental property of any material is its response to a
  localized stress applied at a boundary.  For granular materials
  consisting of hard, cohesionless particles, not even the general
  form of the stress response is known.  Directed force chain networks
  (DFCNs) provide a theoretical framework for addressing this issue,
  and analysis of simplified DFCN models reveal both rich mathematical
  structure and surprising properties.  We review some basic elements
  of DFCN models and present a class of homogeneous solutions for
  cases in which force chains are restricted to lie on a discrete set
  of directions.
\end{abstract}

\section{Introduction}
\label{sec:intro}
Cohesionless granular materials exhibit a range of behavior that has
fascinated physicists, applied mathematicians, and engineers from
Coulomb, in the late $18^{th}$ century, to Schaeffer, in the late
$20^{th}$ and early $21^{st}$.  It is therefore somewhat surprising
that perhaps the simplest question one can ask about a granular
material remains unanswered even now.  Given a box full of sand, if a
marble is placed on top of the sand, how is the weight of the marble
distributed on the bottom of the box?  The problem is not just
quantitative; we do not even know the qualitative form of the pattern
of vertical force on the bottom generated by the marble.  We do not
know, for example, whether the vertical force is largest directly
underneath the marble or has a maximum on a ring. \footnote{A review
  of the many and varied approaches that have been taken to the
  problem of determining stress distributions in noncohesive granular
  materials is beyond the scope of the present paper.  The reader is
  referred to References~[1-13] and the citations therein for
  background material.}  

Experiments suggest that qualitative features of the response may
depend on the way in which the sand was put into the box or on the
geometric and frictional properties of the
grains.\cite{behringer,serero}.  Experiments on 2D layers with depths
up to approximately 12 grain diameters show both two-peak and
single-peak responses, depending on the geometry of the grains and/or
degree of disorder in the packing. \cite{behringer} Experiments on 3D
layers have probed the response in layers as deep as 300 grain
diameters, and single peaks directly beneath the applied load were
observed. \cite{serero}  A theoretical framework that allows a unified
interpretation of these results would be of great interest.

Let us formulate a thought experiment that focuses on the central
features governing the stress pattern at the bottom of the sandbox.
The fundamental problem has to do with how stress is transmitted
through the bulk of the material, so it is natural to eliminate the
side walls from consideration by imagining a slab of sand that is
infinite in horizontal extent.  It is also tempting to eliminate the
effect of the bottom boundary by imagining it to be at infinity, where
all stresses are infinitesimal, and calculating the stress as a
function of depth far from the bottom boundary.  We will see, however,
that in the context of an interesting class of models the bottom
boundary condition plays an essential role.

We can also eliminate from the problem the complicating effects of
gravity.  We consider a slab of sand between two plates that are being
pushed together.  The density of the sand is assumed to be high enough
that the material rigidly resists the force applied by the plates.
The material is then subjected to a localized force created by poking
a needle through the top plate and applying a specified force to it.
Note that both plates and the confining pressure are essential in this
setup.  Without them, some grains at the surfaces of the slab would
likely be subject to outward forces that could not be balanced, so the
material would not support any force at all on the needle.

A third simplification is to assume that the bulk granular material is
isotropic.  That is, we assume that the geometry of the grain packing
does not distinguish any particular direction in space.  While the
local environment of any particular grain is clearly anisotropic, we
assume that this anisotropy, as measured for example by the fabric
tensor which characterizes the degree to which there are preferred
directions for ``bonds'' joining the centers of grains in contact,
vanishes when coarse-grained over sufficiently large volumes.  This is
a strong assumption, as construction history of a real granular
material may well distinguish favored directions for
contacts.\cite{fabric} Nevertheless, there are highly nontrivial
structures to be uncovered even in isotropic models and it is surely
appropriate to understand them before attempting to include the
complications and additional parameters associated with intrinsic
anisotropies.

Finally, for simplicity, we can work with a two-dimensional system.
The top of the slab is a line defined to be at $z=0$ and the bottom is
defined to be $z=d$.  The slab is infinite in the horizontal $x$
direction.  We are interested in determining the response to a
localized force applied at $x=0$ and $z=0$. 

Treating the sand as an ordinary (linear) elastic medium, a standard
treatment \cite{LLelasticity} indicates that $\sigma_{zz}(x,z)$ should
have a single peak centered on $x=0$ with a width that grows linearly
with depth.\footnote{ In sufficiently strongly anisotropic elastic
  materials, it is possible to have two peaks, both of which have
  widths that grow linearly with depth.\cite{obcs}} This treatment
assumes, however, that the stress field is governed by equations
derived using the continuity of a well defined displacement field and
that there is a unique stress associated with each strain
configuration.  Neither of these is true for a granular material
consisting of rigid particles that support frictional forces, so a new
approach is needed.

The problem of deriving macroscopic stress equations from known
microscopic or grain scale physics has proven quite difficult.  An
indication of just how difficult this might be can be found in almost
any experiment or numerical simulation that generates images of the
intensity distribution of stresses on scales of several grain
diameters.\cite{radjai,behringer,howellPG,blair} In such images, one sees
immediately that the stress is concentrated on filamentary structures
called {\it force chains} that support compressive stress.  These
chains appear to be relatively straight on scales of up to 10 grain
diameters or so and give the visual impression of splitting and fusing
at a variety of angles.  The presence of such structures suggests that
passage from the grain scale to the macroscopic stress will involve
two distinct steps: we must understand how grain scale physics favors
the formation of chains, and also how the interactions among chains
determine the macroscopic stress field.  While these two tasks must
ultimately be facets of a unified theory, it may be useful to approach
them separately.

Models of {\it directed force chain networks} (DFCNs) have recently
been proposed to bridge the gap between the scale of individual chains
and the macroscopic stress, leaving open the issue of how grain scale
physics promotes chain formation.\cite{bclo,scs} (This type of model
is also referred to as a \uy\dy-model.\cite{bclo}).  As detailed
below, a ``Boltzmann equation'' governing the densities of chains with
specified strengths and orientations can be obtained, assuming only
that whether a chain splits at a given point is determined only by the
local environment around that point and that environments that lead to
splitting are homogeneously distributed throughout the material.  An
essential feature of this equation is that the unstressed solution is
unstable; perturbations of it lead to divergent responses.  Thus in
order to calculate physically meaningful response functions, it is
necessary to perturb around a pre-stressed state.  The only case for
which response functions have been calculated is the special one in
which all force chains support the same magnitude of stress and lie on
a 6-fold symmetric set of vectors.  In that case, the response
consists of two peaks whose centers diverge linearly with depth but
whose widths grow like the square root of depth.  In other words, the
only solved case of response functions for DFCNs suggests that stress
propagates along characteristic directions determined by the
homogeneously pre-stressed state, in marked contrast to expectations
from standard elasticity theory.\cite{scs} This propagation of stress
is a feature of noisy hyperbolic equations and is therefore called a
``hyperbolic response.''  \cite{cbcw} 

It is worth noting here that hyperbolic response is known to occur in
models of isostatic packings of frictionless grains \cite{witten}
and that simple constitutive relations leading to hyperbolic stress
equilibrium equations have been shown to explain nontrivial features
of experiments on sandpiles and granular columns. \cite{osl} One must
also note, however, that direct measurements of the response seem to
indicate that single-peaked response is associated with strongly
disordered packings, and hyperbolic response occurs only in ordered
systems or at shallow depths. \cite{behringer,serero}

The question immediately arises as to whether the hyperbolic response
of the 6-fold DFCN model is an artifact of the confinement to a
discrete set of directions, or perhaps of the unique property that all
chains in the 6-fold DFCN have the same strength.  To investigate the
robustness of the 6-fold DFCN results, one would like to find other
analytically tractable cases.  Here we address the first step in this
direction -- the identification of models for which stable solutions
can be found for homogeneous macroscopic stresses.

In this paper we present homogeneous solutions to the DFCN Boltzmann
equation for special cases in which chains lie along a discrete set of
directions with 8-fold, or more generally 4N-fold, symmetry in two
dimensions.  Even the homogeneous solutions -- the pre-stressed states
about which we might contemplate calculating response functions --
have nontrivial features.  The actual calculation of the response
functions these solutions is beyond our present scope.

This paper also provides corrections to the presentation fo the general
\eqname\ equation of Ref.~\cite{scs}.  The corrections do not affect any of 
the calculations on discrete models reported in that paper, but do clarify
some conceptual points crucial for future work on continuum models or
models permitting multiple splitting angles.

\newpage
\section{The DFCN model and \Eqname equation}
\label{sec:master}
\subsection{Definitions}
\label{sec:chaindef}
\begin{figure}
\begin{center}
  \epsfxsize=0.8\linewidth \epsfbox{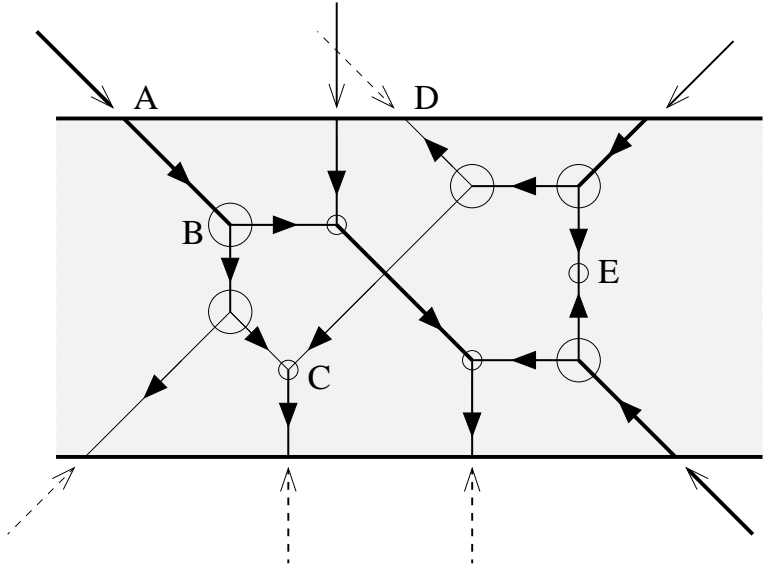}
\caption{Illustration of a directed force chain network.  Line
  thicknesses indicate force chain intensities.  Arrows indicate force
  chain directions.  Large circles mark regions of the granular
  material containing local configurations that require the splitting
  of a force chain entering from the direction pictured.  Small
  circles indicate local configurations where fusion of two incoming
  chains in the directions pictured occurs.  The boundary forces
  applied to the system are shown as solid vectors above the top
  surface and below the bottom surface.  The dashed vectors at the two
  surfaces indicate forces exerted by the top and bottom plates on the
  material inside as a response to the applied forces.
\label{fig:dfcn}}
\end{center}
\end{figure}

A DFCN is a collection of line segments with associated compressive
forces assigned such that force balance is achieved at every vertex.
Figure~\ref{fig:dfcn} shows a simple example that illustrates several
features.  First, each segment is assigned a strength, or force
intensity, that is indicated in the figure by the thickness of the
line.  Every chain is assumed to be under compression, so the force
exerted on a vertex due to a given chain is directed along the chain
toward the vertex, and the vector sum of the forces at any given
vertex must vanish.

Second, each segment is assigned a direction, indicated by the arrow,
which distinguishes its ``beginning'' from its ``end.''  The physical
distinction lies in the local grain configurations at either end of
the chain.  Certain configurations do not allow propagation of a chain
through them in certain orientations and therefore give rise to force
chain splitting.  The relevant configurations of this type for the
network shown in the figure are indicated with large circles.  Other
local configurations would not necessarily require a splitting event,
but do permit two chains to fuse when they intersect.  The
relevant configurations of this type are indicated with small circles.

The directions of chains in the entire network are determined by the
specification of the applied boundary forces.  One must think, for
now, of the force being applied at point A as coming from a needle
being poked through the top plate and held at constant force.  This
ensures that a chain beginning at A must be present.  At point B, a
local configuration exists that requires the chain to end.  To balance
the force at B, two chains must be initiated.  Hence the arrows on
these to chains must point away from B.

By tracing the chains initiated by applied forces at the boundary and
taking into account the type of local environments at the chain
endpoints, each chain direction can be uniquely assigned, with the
exception of rare cases in which fusions happen to occur precisely at
places that might be mistaken for splittings.  When all directions are
assigned, one is likely to find some chains that end, rather than
begin, on a boundary.  The boundary forces required to balance these
forces, shown as dashed arrows in Figure~\ref{fig:dfcn} must be
interpreted as a {\it response} to the applied forces.

Finally, there is the possibility that a single chain can appear to
have inconsistent requirements on its direction, as occurs for the
chain passing through point E in Figure~\ref{fig:dfcn}.  Here two
splitting events have initiated chains along the same direction that
meet head on.  Formally, this corresponds to a fusion event in which
the outgoing chain has zero strength, which thus appears as the
annihilation of two chains, and the circle marking this fusion can be
imagined to lie anywhere along the chain.  The probability of such
configurations is determined by the grain size, or, more precisely, by
the ratio of the grain size to the typical separation between force
chains of opposite directions.  In the following, we will take this
probability to be zero. (But keep in mind that a nonzero probability
can be invoked to cut off the divergence in the density of weak force
chains that arises in the 8-fold model.)

\begin{figure}
\begin{center}
  \epsfxsize=\linewidth \epsfbox{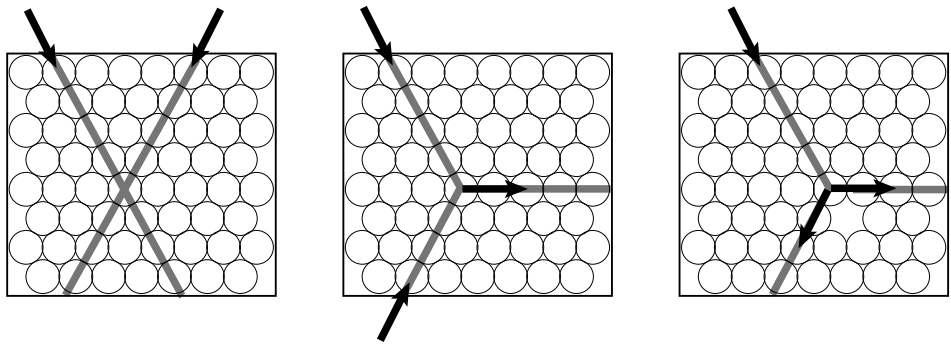}
\caption{Illustration of force chains.  (a) Two force chains
  initiated at the top surface cross without interacting.  (b) When
  one of the force chains is initiated at the bottom, a fusion becomes
  possible.  (c) A force chain initiated at the top that splits upon
  encountering a defect.
\label{fig:forcechains}}
\end{center}
\end{figure}

Figure~\ref{fig:forcechains} further illustrates the difference
between chain splitting and chain fusion in the context of a hexagonal
array of grains.  In panel (a), two forces are applied at the top
boundary and the forces propagate along chains, crossing at the
central grain.  In panel (b), a similar situation is shown, but this
time one of the chains is assumed to be specified by fixing its
position at the bottom boundary instead of the top.  In this case, a
fusion of the two chains at the central grain is permitted (though not
required) and the resulting configuration may be different.
Panel (c) shows a local configuration (a missing disk) that requires
the splitting of an incoming chain.

Thus each chain is characterized by an intensity, $f$, and a direction
$\nh$, or, equivalently, a vector $\fv$.  Force balance at a vertex
requires that $\sum\fv$ for incoming chains equals $\sum\fv$ for
outgoing chains.  Positive values of $f$ correspond to compressive
stress along a chain. negative values to tensile stress.  To model
noncohesive materials, which do not support tensile stress, we take
$f$ to be always positive definite.  Thus there is no ambiguity
introduced by using $\fv$ to denote the pair $(f,\nh)$.  When
convenient, we will use the angular variable $\theta$ to indicate the
direction of $\nh$.  We will use the term ``$\fv$-chain'' to refer to
a chain with the given strength and direction.

\subsection{The \Eqname\ equation}
{\em The following discussion supersedes the treatment given in
  Ref.~\cite{scs}.}  In that paper, mistaken reasoning was used in
writing the continuum equation.  The mistake does not affect the bulk
of that paper, since the forms obtained there for networks of chains
confined to a discrete set of directions were the same as those
derived below.  Nevertheless, the corrections made here may be
important for future work on continuum models and are necessary for
conceptual clarity.  Note especially that the definition of $P$ has
different dimensions here from that used in Ref.~\cite{scs}.  The
theory is developed here explicitly for the case of two dimensions.
Generalization to higher dimensions is straightforward.

Let $P(f,\theta,\rv)$ represent the density (in an ensemble average)
of force chains of intensity $f$ and direction $\theta$ passing
through the spatial point $\rv$.  In other words,
\begin{equation}
\int_{f}^{f+\delta f}\!\!\!\!\!f'\,df'
\int_{\theta}^{\theta + \delta\theta}\!\!\!\!\! d\theta'\;|\nh\!\cdot\!\uh|P(f',\theta',\rv)
\end{equation} 
is the number of chains with intensity between $f$ and $f+\delta f$
that cross a unit length line segment passing through $\rv$ and
perpendicular to $\uh$.  With this definition, $P(\fv)$ is defined
with respect to a uniform measure in the 2D space of possible forces.
Given $P(\fv,\rv)$, the local macroscopic stress tensor is given by
\cite{bclo}
\begin{equation}\label{eqn:stresstensor}
\sigma_{\alpha\beta} = \int_{0}^{\infty}\!\!\!\!f df \int_{-\pi}^{\pi}\!\!d\theta\;n_{\alpha}n_{\beta} f P(f,\theta,\rv).
\end{equation}

We wish to construct an equation that describes the ensemble average
of the spatial variations in the chain densities for a system subject
to specified boundary conditions.  To do so, we consider the
probability that a chain $\fv$ will exist at a point $\rv + \epsilon
\nh$, assuming we know $P(f,\nh,\rv)$.

A given material will be characterized by a two scalar parameters,
$\lambda$ and $Y$, and two angular functions, $\phi_s$ and $\phi_f$.
\begin{itemize}
\item $\lambda$ is the average distance in any specified direction
  between the point where a chain begins and the nearest point that
  will cause it to split; i.e., the mean length chains would have if
  there were no fusions.
\item $\phi_s(\fv_1\,|\,\fv_2,\fv_3)$ is probability that a given
  splitting of a $\fv_1$-chain results in chains with strengths and
  directions given by $\fv_2$ and $\fv_3$, normalized to unity in the
  sense defined below.
\item $\phi_f(\fv_1\,|\,\fv_2,\fv_3)$ is relative probability that a
  $\fv_2$-chain and a $\fv_3$-chain will fuse when they intersect and
  thereby form a $\fv_1$-chain, normalized to unity in the sense
  defined below.
\item $Y$ is an overall efficiency with which two intersecting chains
  will fuse.  That is, the total probability that two intersecting
  chains will fuse is $Y\phi_f$
\end{itemize}
$\phi_s$ and $\phi_f$ must both include delta functions that enforce
force balance at each intersection and Heaviside functions that
guarantee all forces are compressive.  

We neglect, for now, the finite size of individual grains.  That is,
we assume that $\lambda$ is large compared to a grain diameter and
treat the granular packing underlying the force chain network as a
continuous medium.  The equation governing the spatial variation of
$P$ as follows (for notational convenience, we have dropped
the $\rv$ from the argument of all of the $P$'s):
\begin{eqnarray}\label{eqn:master}
(\nh\cdot\nabla) P(\fv) & = & 
        -\frac{1}{\lambda}\int\! f'df'\,d\theta'\,f''df''\,d\theta''\;\left|\sin(\theta'-\theta'')\right|\phi_{s}(\fv\,|\,\fv',\fv'')P(\fv) \\
\ & \ & +\frac{2}{\lambda}\int\! f'df'\,d\theta'\,f''df''\,d\theta''\;\left|\sin(\theta-\theta'')\right|\phi_{s}(\fv'\,|\,\fv,\fv'')P(\fv') \nonumber \\
\ & - & 2\, Y\int\! f'df'\,d\theta'\,f''df''\,d\theta''\;\left|\sin(\theta-\theta'')\right|\phi_{f}(\fv'\,|\,\fv,\fv'')P(\fv)P(\fv'') \nonumber\\
\ & + & Y\int\! f'df'\,d\theta'\,f''df''\,d\theta''\;\left|\sin(\theta'-\theta'')\right|\phi_{f}(\fv\,|\,\fv',\fv'')P(\fv')P(\fv'') \nonumber
\end{eqnarray} 
Each term on the right hand side represents a type of event that can
alter the density $P(\fv)$.  We discuss each in turn:
\begin{itemize}
\item The first term represents the loss of $\fv$-chains due to
  splitting.  The $\sin$ factor is exhibited explicitly for future
  convenience.  Here $P(\fv)$ may be taken out of the integral.  To
  ensure that $\lambda$ is the mean distance between a chain will
  propagate before splitting (in the absence of interactions with
  other chains), we must therefore have
  \begin{equation}\label{eqn:phisnorm}
  \int\! f'df'\,d\theta'\,f''df''\,d\theta''\;\left|\sin(\theta'-\theta'')\right|\phi_{s}(\fv\,|\,\fv',\fv'') = 1.
  \end{equation}
  Note that $\phi_s$ has dimensions of $1/force^4$.
\item The second term represents the creation of $\fv$-chains due to
  the splitting of chains in other directions.  The splitting function
  here must be the same as in the first term, but with arguments
  exchanged, since the rate at which chains chains appear due to
  splitting must match the rate at which the parent chains split.  The
  factor of $2$ counts the identical integral arising from the
  exchange of the prime and double-prime labels.  
\item The third term, which is nonlinear in $P$, represents the loss
  of $\fv$-chains due to their fusions with $\fv''$-chains (or
  $\fv'$-chains, hence the factor of $2$).  The quantity
  $\left|\sin(\theta-\theta'')\right|P(\fv)P(\fv'')$ is the density of
  intersections of $\fv$ and $\fv''$-chains, and $\phi_f$ is specifies
  the probability of fusion when two such chains meet.  Here again
  $P(\fv)$ can be taken out of the integral.  We choose to normalize
  $\phi_f$ such that the remaining integral has a maximum value of
  unity over the set of $P(\fv'')$'s of the form $P(\fv'') =
  \delta(\fv''-\fv_0)$, for all values of $\fv_0$.  In other words, we
  normalize $\phi_f$ according to the type of intersection most likely
  to produce a fusion if all chain densities were identical.  In
  equation form:
  \begin{equation}
  \max\left[\int\! f'df'\,d\theta'\,f''df''\,d\theta''\;\phi_{f}(\fv'\,|\,\fv,\fv'')\delta(\fv''-\fv_0)\right] = 1,
  \end{equation}
  or, equivalently, 
  \begin{equation}\label{eqn:phifnorm}
  \max\left[\int\! f'df'\,d\theta'\;\phi_{f}(\fv'\,|\,\fv,\fv_0)\right] = 1,
  \end{equation}
  where the maximization is over all possible values of $\fv$ and
  $\fv_0$.  The parameter $Y$ then provides an absolute measure of the
  fusion efficiency.  $\phi_f$ has dimensions $1/force^2$, in contrast
  to $\phi_s$.
\item The fourth term represents the creation of $\fv$-chains due to
  the fusion of chains in other directions.  For consistency, the
  fusion function must be the same as in the third term.
\end{itemize}

For materials composed of perfectly rigid grains, a rescaling of all
of the forces in a given DFCN yields another perfectly acceptable
network.  Hence, Eq.~(\ref{eqn:master}) must be invariant under a
rescaling of all the force intensities.  This is verified by
straightforward dimension counting (unlike the form of the Boltzmann
equation suggested in Ref.~\cite{scs}).  

Consider further the general forms of $\phi_{s}$ and $\phi_{f}$.
For an isotropic material, $\phi$ can depend only on the differences between angles.
The general form is
\begin{equation}
\phi_{a}(\fv\,|\,\fv',\fv'') =
F_{a}(f,f',f'')\delta(\fv' + \fv'' - \fv)
\Theta(f)\Theta(f')\Theta(f'')\psi_{a}(\theta'-\theta,\theta''-\theta),
\end{equation}
where $F_{a}$ is a combination of the force intensities that provides
the correct dimensions for $\phi_{a}$, and $\psi_{a}$ is a symmetric
function of its arguments.  (One might imagine functions of the ratios
of force intensities multiplying the arguments of $\psi_{a}$, but
these will always reduce to functions of the angles alone due to the
$\delta$ function.)

In the case of fusions, dimension counting in
Eq.~(\ref{eqn:phisnorm}) implies $F_f$ is dimensionless and hence
equal to unity, up to a constant that can be absorbed in $Y$.  In the
case of splitting, on the other hand, dimension counting in
Eq.~(\ref{eqn:phisnorm}) implies $F_s$ has dimension $1/force^2$.
The relations between the $f$'s enforced by the $\delta$ function
guarantee that any combination of $f$, $f'$, and $f''$ with the right
dimensions is equally valid; the differences between them can simply
be absorbed into $\psi_{s}$.  The natural choice is $F_{s} =
1/(f'f'')$.  With this choice a constant $\psi_{s}=1/\pi^2$
corresponds to an equal probability for every possible splitting
configuration.

It is clear that $\lambda$ can be scaled to unity without loss of
generality by choice of the unit of length.  Though $Y$ is a
dimensionless parameter with physical significance, from a
mathematical point of view it plays a trivial role in
Eq.~(\ref{eqn:master}).  To solve for $P$ for any given value of $Y$,
we solve the case $Y=1$, then simply divide all $P$'s by $Y$.  From
here on, we take $\lambda=1$ and $Y=1$.

\subsection{Specialization to discrete directions}
\begin{figure}
\begin{center}
  \epsfxsize=0.7\linewidth \epsfbox{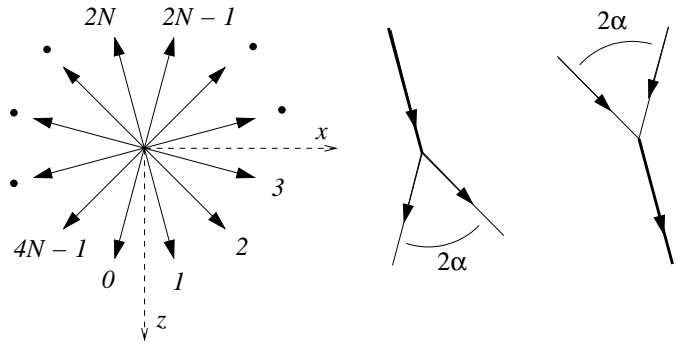}
\caption{Star of chain directions and allowed vertices for solvable models.
  The star has $4N$-fold symmetry.  The vertices show a splitting and a fusion
  with angle $\alpha = \pi/(2N)$.
\label{fig:vertices}}
\end{center}
\end{figure}

We now make two simplifying assumptions to arrive at a set of ordinary
differential equations that permits analytical solution.  First, we
assume that $\phi_s$ and $\phi_f$ are nonzero only for vertices of the
forms shown in Figure~\ref{fig:vertices}.  This means that all chains
lie in the discrete set of directions $\theta_n =
(n-\frac{1}{2})\pi/(2N)$ (measured from the positive $z$ direction,
which is downward), and that the strong force at a given vertex is
always related to two weak ones by a factor of $\xi = 2\cos(\alpha)$.
Defining $f_m\equiv f_0\xi^m$, the continuous function $P(f,\theta)$
becomes a set of discrete functions $P(f_m,\theta_n)$, which will be
denoted $P_{n,m}\delta(\fv-f_m\nh_n)$.  In this notation, it is always
assumed that $n$ is taken modulo $4N$.  Note that $P_{n,m}$ is simply
a number per unit length; the dimensions of force are taken care of by
the two-dimensional delta function.  We will treat the case $\alpha=
\pi/(2N)$ explicitly here.  Generalization to any $\alpha$ that is an
integer multiple of this is straightforward.  

This first assumption corresponds to the following forms for the angular
parts of the splitting and fusion functions:
\begin{eqnarray}
\psi_{s}(\varphi',\varphi'')  & = & \frac{1}{2}\left[
  \delta(\varphi'-\alpha)\delta(\varphi''+\alpha) + \delta(\varphi'+\alpha)\delta(\varphi''-\alpha)\right].\\
\psi_{f}(\varphi',\varphi'')  & = & \left\{
    \begin{array}{ll}
    1 & {\rm if\ } \varphi'= -\varphi''=\alpha {\rm\ or\ } \varphi'= -\varphi''= -\alpha \\
    0 & {\rm otherwise.}
    \end{array}\right.
\end{eqnarray}
The difference in character between $\phi_s$ and $\phi_f$ stems from
the fact that $\phi_f$ appears in integrals that are quadratic in $P$.
When $P$ is a sum of two-dimensional delta functions, the fusion
integrals have two more delta function factors than the splitting
integrals.  The same can be seen in the normalization conditions of
Equations~(\ref{eqn:phisnorm}) and~Equations~(\ref{eqn:phifnorm}).

Second, we assume that the boundary conditions of interest are uniform
across the top and bottom of the slab.  Translational symmetry in the
$x$-direction then dictates that all horizontal gradients of $P_{n,m}$
vanish.

Inserting these assumptions into Eq.~(\ref{eqn:master}) and
integrating both sides over a small volume of force space in the
vicinity of $f_m\nh_n$, we obtain the following ordinary differential
equations for the $P_{n,m}$'s:
\begin{eqnarray}
\label{eqn:discrete}
(\cos\theta_n)\,\partial_z P_{n,m} & = & - P_{n,m} + P_{n+1,m+1} + P_{n-1,m+1} \\
\ & \ & + P_{n+1,m-1} P_{n-1,m-1} - P_{n,m} P_{n-2,m} - P_{n,m} P_{n+2,m}.  \nonumber
\end{eqnarray}
The result is straightforward, but one must be careful to account for
all the trigonometric factors arising from integrating over the delta
functions, including the integral on both sides that is required to
isolate $P_{n,m}$.

The different terms on the right hand side of
Eq.~(\ref{eqn:discrete}) correspond to the processes described
above that can create or destroy chains.  In the present case, the
first term represents the decay of $P_{n,m}$ along the direction
$\theta_n$ due to the splitting of an existing $(n,m)$ chain.  The
next two (linear) terms account for the addition to $P_{n,m}$ due to
the splitting of other chains in adjacent orientations.  Note that if
a chain is to split and create a contribution to $P_{n,m}$, which must
have strength $f_0 \xi^m$, it must begin with strength $f_0 \xi^{m+1}$.
The first nonlinear term accounts for events in which two chains of
strength $f_0 \xi^{m-1}$ fuse, and the last two nonlinear terms to
events in which chains contributing to $P_{n,m}$ are deflected due to
fusions with other chains.

In Reference~\cite{scs}, Eq.~(\ref{eqn:discrete}) was studied in
detail for the case of 6-fold symmetry.  In that case we have $\xi =
1$, so the hierarchy of equations indexed by $m$ collapses.  In
addition, $\cos\theta_2 = \cos\theta_5=0$, so the entire system
reduces to four ODEs and two algebraic relations.  If we assume mirror
symmetry about the vertical axis, these are reduced to two
equation that can be solved completely.  

In order to observe the variations of $P$ with force strength, we must
consider symmetries other than 6-fold.  To avoid technical
complications arising from the presence of strictly horizontal chains
having $\cos\theta = 0$, we consider only the cases of $4N$-fold
symmetry for positive integers $N$.  The simplest case is that of
8-fold symmetry, and example of which is illustrated in
Figure~\ref{fig:dfcn}.

Note that the discreteness of the possible orientations of force
chains does {\em not} imply any discreteness in the possible positions
of the splitting or fusion vertices.  Note also that the system
remains isotropic in the sense that Eq.~\ref{eqn:discrete}, including
all coefficients, is identical for all $n$, except for the geometric
factor on the left-hand side that results from restricting attention
to variations in $z$ and not $x$.

\section{Homogeneous networks for symmetric splittings and fusions}
\label{sec:discrete}
General solutions of Eq.~(\ref{eqn:discrete}) are not yet known.
We study here the important special case of homogeneous solutions;
i.e., the fixed points, for which all gradients vanish.  These
solutions are both nontrivial and crucial for understanding the
response function.  Consider the trajectory of the vector $P_{n,m}$ as
$z$ is varied.  As detailed in Reference~\cite{scs} for the 6-fold
case, trajectories that do not pass very close to a stable (or
marginally stable) fixed point inevitably lead to divergences at
$z>>\lambda$ that are reflected in unphysical negative values of
some $P_{n,m}$.  Trajectories that avoid these divergences do so by
staying in the vicinity of a fixed point over most of the range of
$z$.  As also noted in Reference~\cite{scs}, the trivial fixed point
$P_{n,m}=0$ is unstable and hence cannot serve as a reference state
for a linear response theory.  Thus the response function we
ultimately seek will be dominated by the DFCN structure corresponding
to some nontrivial fixed point.

\subsection{Isotropic solutions}  
It is instructive to begin with a search for isotropic solutions;
i.e., solutions for which $P_{n,m} = P_{m}$ for all $n$.  The fixed
point equation derived from Eq.~(\ref{eqn:discrete}) reads
\begin{equation}\label{eqn:Pisotropic}
-P_{m} + 2 P_{m+1} + P_{m-1}^2 - 2 P_{m}^2 = 0.
\end{equation}
To simplify this recursion relation, define 
\begin{equation} \label{eqn:dm}
d_m = P_m - P_{m-1}^2.
\end{equation}
Eq.~(\ref{eqn:Pisotropic}) says $2 d_{m+1} - d_m = 0$, which
implies
\begin{equation}
d_m = d_0 2^{-m}.
\end{equation}

For $d_0\neq 0$, Eq.~(\ref{eqn:dm}) gives
\begin{equation}
P_{m+1} =  P_m^2 + d_0 2^{-m}.
\end{equation}
For $d_0 < 0$, this recursion relation either produces a negative
numbers or a divergence at large $m$, neither of which is physically
acceptable.  If $P_0$ is sufficiently small, the repeated squaring
drives $P_m$ toward zero at least as fast as $p^{-2^m}$ with $p>1$.
The negative contribution from the $d_0$ term can only speed up this
decay as long as $P_m$ is positive.  But this means that $P_m$ decays
faster than $2^{-m}$, so the $d_0$ term eventually makes $P_m$
negative.  Now if $P_0$ is large enough to avoid $P_m$ decaying toward
$0$, repetitive squaring causes it to diverge like $p^{2^m}$ with
$p>1$; the $d_0$ term becomes irrelevant at large $m$.

For $d_0>0$, on the other hand, problems arise for large, negative $m$.
We write $P_{m-1} = \sqrt{P_m - d_0 2^{-m}}$ and see that complex
values will be generated unless $P_m$ is always greater than $d_0
2^{-m}$.  But this is impossible because $P_{m-1}$ is strictly less
than $P_m$. 

Thus we are left with $d_0 = 0$ as the only physically relevant case.
When $d_0 = 0$, Eq.~(\ref{eqn:discrete}) is satisfied in a special
way: both $-P_{m} + P_{m-1}^2 = 0$ and $2 P_{m+1} - 2 P_{m}^2 = 0$ are
satisfied simultaneously.  These two relations are actually identical
up to a shift in $m$, and admit a one-parameter family of solutions
\begin{equation}
P_m = p^{-2^m},
\end{equation} 
where we must have $p>1$ to avoid divergence at large $m$.  Despite
its apparent simplicity, it is instructive to examine this solution in
a bit more detail.  We consider first the behavior at large negative
$m$, then the behavior at large positive $m$.

Note that $P_m \rightarrow 1$ for large negative $m$.  This means that
the densities of all chains of intensity $f_0 \xi^{-m}$ for large $m$
are the same.  In other words, the total density of force chains,
$\sum_m P_m$ diverges.  Though this may appear troublesome, it does
{\it not} yield a divergence in the pressure: $\sum_m P_m f_m$ remains
perfectly finite.  The divergence in chain density is an artifact 
of the restriction to a fusion function $\phi_f$ that does not allow
weak chains to fuse with stronger ones.
Removal of this artifact requires working with a continuous
distribution of force chain directions and intensities and it beyond
the scope of this work.

\begin{figure}
\begin{center}
  \epsfxsize=0.7\linewidth \epsfbox{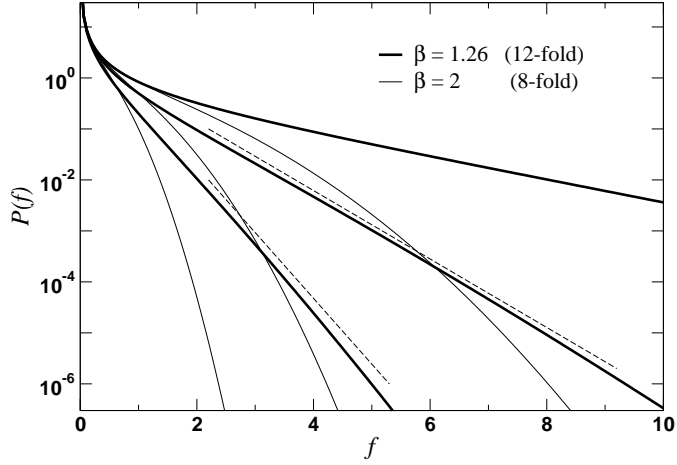}
\caption{Single contact force distributions for 8-fold and 12-fold models.  
  For each case, curves are shown for $p=1.2$, $2$, and $5$.  The
  straight, dashed lines are guides to the eye to emphasize the fact
  that the curves look very close to ordinary exponential decays.
\label{fig:poff}}
\end{center}
\end{figure}

For large positive $m$, we can compute the probability distribution
for contact forces, a quantity that has been shown in a variety of
experiments and numerical simulations to decay exponentially for large
forces.  For the homogeneous, isotropic solution, we have
\begin{equation}
{\mathcal P}(f_m) = P_m / (f_m - f_{m-1}),
\end{equation}
where $f_m = f_0 \xi^m$ and $P_m = p^{2^m}$.  The denominator in this
expression is just the spacing between points with successive $m$'s,
the required conversion factor from the discrete density $P_m$ to the
density per unit force.  This can be rewritten as
\begin{equation}
{\mathcal P}(f)\simeq p^{-(f/f_0)^\beta}/f,\quad{\rm with}\quad \beta=\frac{\log 2}{\log \xi}.
\end{equation}
Recall that $p>1$ and $\xi<2$, so $\beta>1$ and ${\mathcal P}$ decays
faster than exponentially for large $f$.  For the 8-fold case, we have 
$\beta = 2$ and hence a Gaussian decay.
But smaller splitting angles give $\xi$ closer to 2, and for a splitting
angle as large as $2\alpha = 60^{\circ}$, we have $\beta \approx 1.26$.
In this case, the full distribution (including the $1/f$ term) can
look surprisingly like a simple exponential over several
decades, as shown in Figure~\ref{fig:poff}.

Experience with the 6-fold model suggests that the isotropic fixed
points of Eq.~(\ref{eqn:discrete}) are non-generic.  In the 6-fold
model, linearization of the theory in the vicinity of a fixed point
generally leads to hyperbolic response functions with peaks that are
sharper for more strongly anisotropic fixed points.  The isotropic
case is pathological in that the peak width is divergent\cite{scs} and
additional cancellations conspire to produce single peak, but one that
obeys different scaling laws from the predictions of elliptic
models.\cite{scunpublished} It is therefore important to consider
solutions of Eq.~(\ref{eqn:discrete}) that are not constrained to be
isotropic.

\subsection{Anisotropic solutions}
To find fixed points of Eq.~(\ref{eqn:discrete}), we begin with
the following ansatz :
\begin{equation}
P_{n,m} = p^{q_m + x_n \zeta^{m}},
\end{equation}
where $p$ is a real parameter and $q_m$, $x_n$, and $\zeta$ are to be
determined.  (This guess combines features of the isotropic solution
above and the 6-fold solutions discussed in Reference~\cite{scs}.)
Substituting into Eq.~(\ref{eqn:discrete}), we have
\begin{eqnarray}\label{eqn:pqx}
0 & = & -\; p^{q_m + x_n \zeta^{m}} + p^{q_{m+1} + x_{n-1} \zeta^{m+1}} + p^{q_{m+1} + x_{n+1} \zeta^{m+1}} \\
\ & \ & +\; p^{2q_{m-1} + (x_{n+1}+x_{n-1}) \zeta^{m-1}} - p^{2q_m + (x_n+x_{n-2})\zeta^{m}} - p^{2q_m + (x_n+x_{n+2})\zeta^{m}}. \nonumber
\end{eqnarray}
Taking a cue from the isotropic case, we consider the possibility that
the cancellation required by this equation occurs term by term in the
following way:\footnote{Unlike in the isotropic case, we do not prove
  here that all physically plausible solutions {\em must} have this
  form.}
\begin{eqnarray}
p^{q_m + x_n \zeta^{m}} & = & p^{2q_{m-1} + (x_{n+1}+x_{n-1}) \zeta^{m-1}}; \label{eqn:pqx1} \\
p^{q_{m+1} + x_{n-1} \zeta^{m+1}} & = & p^{2q_m + (x_n+x_{n-2})\zeta^{m}};  \label{eqn:pqx2} \\
p^{q_{m+1} + x_{n+1} \zeta^{m+1}} & = & p^{2q_m + (x_n+x_{n+2})\zeta^{m}}.  \label{eqn:pqx3}
\end{eqnarray}
Note that all three of these equations are identical up to shifts of
$n$ and $m$.  Since the $q$ terms are independent of $n$, we obtain
two recursion relations:
\begin{eqnarray}
q_{m+1} &  = & 2q_{m};             \label{eqn:qm} \\
x_{n} \zeta & = & x_{n-1}+x_{n+1}. \label{eqn:xnm} \\
\end{eqnarray}
Eq.~(\ref{eqn:qm}) implies 
\begin{equation}
q_m = -q_0 2^m,
\end{equation}
where the negative sign defines a convention for the sign of $q_0$.
Eq.~(\ref{eqn:xnm}) is an eigenvalue equation for a $4N\times 4N$
matrix $\Am$.  The eigenvalues $\zeta_j$ and eigenvectors $y^{(j)}_n$
of $\Am$ are easily found:
\begin{equation}
\zeta_j = 2\cos\left(\frac{j \pi}{2N}\right); \quad\quad
y^{(j)}_n = \left\{\begin{array}{ll}
  \cos\left(\frac{j n \pi}{2N}\right) & j=0,1,\ldots 2N \\
  \sin\left(\frac{j n \pi}{2N}\right) & j=2N+1, \ldots 4N-1
             \end{array}\right.
\end{equation}
Thus for each of the $4N$ eigenvalues and eigenvectors, we have what
appears to be a three-parameter family of fixed point solutions of
Eq.~(\ref{eqn:discrete}) of the form
\begin{equation}\label{eqn:solutions}
P_{n,m} = p^{-q_0 2^m + x_0 y^{(j)}_n (\zeta_j)^m},
\end{equation}
the free parameters being $p$, $q_0$, and $x_0$.  Recall that the
force intensity associated with the chains whose density is $P_{n,m}$
is $f_m = f_0 \xi^m$ and note that $\xi = \zeta_1$.  Note also that
for $j=0$, we have $\zeta=2$, so that the $x_0$ and $q_0$ terms can be
combined.  This leaves us with precisely the same isotropic solution
discussed above.

\begin{figure}
\begin{center}
  \epsfxsize=0.95\linewidth \epsfbox{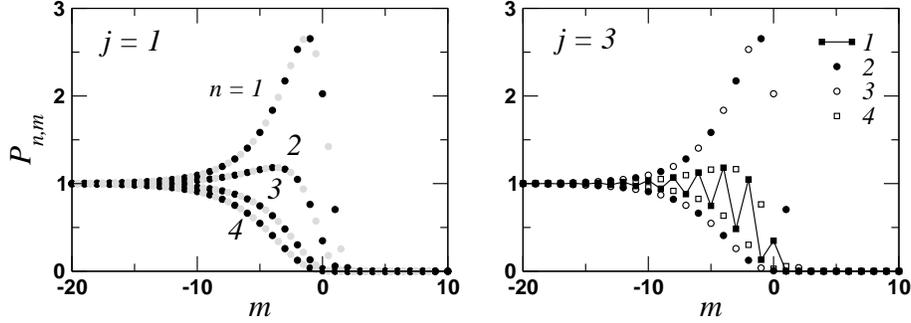}
  \caption{Anisotropic solutions for 8-fold case.  The solutions
    pictured have reflection symmetry about the vertical; i.e.,
    $P_{n,9-m}=P_{n,m}$.  Left: Two $j=1$ ``dipole-like'' solutions.
    Black circles show a solution corresponding to $p=10$,
    $x_0=q_0=1$. Grey circles show a solution for $x_0=1.7$, with $p$
    adjusted and $m$ shifted to show that it belongs in the same
    family as the $x_0=1$ solution.  Right: A solution with a higher
    order anisotropy, $j=3$.  The oscillation of $P_{n,m}$ with $m$ is
    apparent.  The solid line is a guide to the eye for following the
    oscillations of $P_{1,m}$.
\label{fig:solutions}}
\end{center}
\end{figure}

Eq.~(\ref{eqn:solutions}) requires further examination to
determine the range of different structures it represents.  First,
each three-parameter family is really only a one-parameter family of
distinct physical networks.  There are ways of rescaling the
parameters that have no effect on the physical system begin described:
\begin{eqnarray}
(q_0, x_0) & \rightarrow & (2^\nu q_0, \zeta^\nu x_0)\quad {\rm and} \label{eqn:rescalingf}\\
(p, q_0, x_0) & \rightarrow & (p^{q_0}, 1, x_0/q_0).
\end{eqnarray}
The first of these corresponds to a redefinition of the force scale,
$f_0 \rightarrow f_0 \xi^\nu$, which cannot affect the physics.
Strictly speaking, the system is exactly invariant under this
transformation only if $\nu$ is an integer.  Arbitrary values of
$\nu$, however, merely shift the positions of the discrete set of
$P_{n,m}$'s along a single smooth curve for each $n$, with integral
$\nu$'s being the values that shift the entire set into itself.  For
our purposes, the differences between solutions related by non-integer
$\nu$ are unimportant details.  (See Figure~\ref{fig:solutions}.)  The
second transformation leads to a mathematically identical solution.
(The case $q_0=0$ is not physically relevant, as it leads to
divergence of the total stress from large $m$ force chains.)  By
performing the first transformation with $\nu =
\log(x_0/q_0)/\log(2/\zeta)$, followed by the second, we can adjust
$q_0$ and $x_0$ both to unity.

Second, note that we {\it cannot} take arbitrary linear combinations
of the solutions to Eq.~(\ref{eqn:xnm}), since they would not
satisfy the nonlinear Eq.~(\ref{eqn:discrete}).  We can, however,
take linear combinations of eigenvectors that have degenerate
eigenvalues.  Since $\zeta_{4N-j} = \zeta_j$ for $1\leq j\leq 2N-1$ we
can generate solutions from arbitrary linear combinations of the
degenerate pair of eigenvectors $\yv^{(j)}$ and $\yv^{(4N-j)}$.  Thus
$j=0$ and $j=2N$ each provide a single continuous family of solutions
parameterized by $p$, while all other $j$'s come in pairs that provide
families parameterized by $p$ and by the relative weights of the two
$\yv^{(j)}$'s.
  
Consider now the family spanned by $\yv^{(j)}$ and $\yv^{(4N-j)}$, for
some specific $j$.  By forming linear combinations, we can construct
basis vectors $\ev_j$ that are symmetric or antisymmetric under
reflection through a vertical axis.  The symmetric combination will
give solutions for which $P_{n,m} = P_{4N+1-n,m}$, as might be
expected if the average confining force on the plates is purely
normal.  The antisymmetric combination will give solutions that break
this symmetry, though they do {\em not} yield any negative values of
$P_{n,m}$.  These might be important for describing systems subjected
to shear as well as compression.  The explicit forms of the symmetric
combinations (including the trivially symmetric isotropic case) can
be expressed as
\begin{equation}
e^{(j)}_n  = 
 \left\{\begin{array}{ll}
  1                                     & j=0  \\
  \sqrt{2} \cos(j\theta_n) & j=1,\ldots,2N-1,
        \end{array}\right.
\end{equation}
where the normalization is $||\ev_j||=4N$ for all $j$. 

The
antisymmetric combinations, normalized to $4N$, are given by
\begin{equation}
e^{(j)}_n  = 
 \left\{\begin{array}{ll}
  (-1)^n                                     & j=2N  \\
  \sqrt{2} \sin(j\theta_n) & j=2N+1,\ldots,4N-1,
        \end{array}\right.
\end{equation}
In the following, we focus on the symmetric cases only.

The $\ev^{(0)}$ solution is the isotropic case discussed above.  The
$\ev^{(1)}$ solution describes a stress configuration with a dipole-like
anisotropy.  At each generation $m$, the angular variation $P_{n,m}$
has a single maximum (for $p>1$) at $n=1$ and a minimum $n=2N$.  There
are more downward-pointing than upward-pointing chains for each
undirected orientation.  Moreover, because $\zeta_1$ is positive, the
decay of $P_{n,m}$ with $m$ is monotonic both at large $m$, where it
decays to $0$, and at large negative $m$, where it decays to $1$.
Figure~\ref{fig:solutions} shows the form of $P_{n,m}$ for this case.
Note that the term ``dipole-like'' applies here to the $P$'s.  Since
the stress field does not distinguish between force chains in opposite
directions, the anisotropy in the stress is ``quadrupole-like'' --
stronger in the vertical directions than in the horizontal.

\begin{figure}
\begin{center}
  \epsfxsize=0.95\linewidth \epsfbox{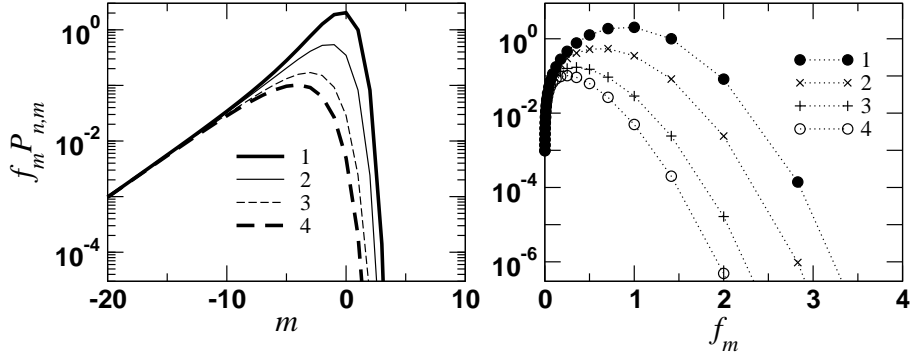}
  \caption{Stresses for an anisotropic solution.  The solution
    pictured is the same as the one on the left in
    Figure~\ref{fig:solutions}.  Here we show the compressive stress
    associated with a given chain direction rather than just the chain
    density, plotted vs.\ $m$ and $f_m$.  The dotted lines on the
    right are just guides to the eye.
\label{fig:stresssoln}}
\end{center}
\end{figure}

A striking feature of the solutions is the occurrence of peaks for
some $n$, which become more pronounced for larger $p$.  Peaks are
evident for all $n$ when the stress, $f_m P_{n,m}$, is plotted rather
than the chain density.  (See Figure~\ref{fig:stresssoln}.)
Differentiating Eq.~(\ref{eqn:solutions}) with $q_0=x_0=1$, we
find that the peak in $P_{n,m}$ occurs at
\begin{equation}\label{peakm}
m = \frac{\log y^{(j)}_n + \log\log|\zeta_n| - \log\log 2}{\log 2 - \log|\zeta_n|},
\end{equation}
which is independent of $p$.  (The peak in $f_m P_{n,m}$ does depend
weakly on $p$.)  The force scale represented by this peak must have
its origin in the boundary conditions, since a
linear rescaling of all forces in the bulk has no effect on the DFCN
structure.  Recall that part of the process required to collapse the
solutions onto the one-parameter family indexed by $p$ was to rescale
$f_0$, which corresponds to a rescaling of the strengths of the force
chains injected at the boundaries.  There are subtleties lurking here,
however, as will be discussed further in
Section~\ref{sec:boundaryconditions}.

A full calculation of the single contact force distribution ${\mathcal
  P}(f)$ is slightly more complicated than for the isotropic case, as
it requires a sum over the different chain orientations.  It is not
hard to extract the dominant term at large $f$, however.  For the
symmetric $j=1$ case with $q_0 = x_0 = 1$,
Eq.~(\ref{eqn:solutions}) can be written as
\begin{equation}\label{eqn:solution1f}
P_{n,m} = p^{-f_m^\beta + e^{(1)}_n f_m^\gamma},
\end{equation}
where $\beta = \log 2/\log\xi$ as above, and $\gamma =
\log\zeta_1/\log\xi = 1$.  Since $\beta>1$, the dominant term at large $f$
will be $P_{1,m} = p^{-f_m^\beta +  f}$.

It is important to note that the anisotropy of the solution is {\em
not} a consequence of any material anisotropy.  The bulk properties of
the material we are describing are perfectly isotropic. The boundary 
conditions on the system will generally not be isotropic, however, so 
there is no reason to expect that the pre-stressed state about which the
response function should ultimately be calculated should have isotropic
stress.  The generic situation described by the discrete DFCN theory 
presented here is one in which an intrinsically isotropic material is
subjected to loading forces that generate anisotropic distributions of
stress chains.  Because different degrees of anisotropy correspond to   
different fixed points of the full nonlinear theory and hence to
different linearizations, the pre-stressed state may have a profound
influence on the response function.  This is in contrast to an ordinary 
elastic medium, in which the response function is not affected by an
anisotropic background stress.

The solutions for higher $j$ have features that seem a bit unlikely
from a physical standpoint, though not obviously unacceptable.  For
$2\leq j < N$, we have higher orders of anisotropy, with $P_{n,m}$
exhibiting additional maxima and minima as a function of $n$ at each
$m$, but still with monotonic decay at large $m$.  For $N< j \leq 2N$,
the eigenvalues are negative, which causes $\zeta^{m}$ and hence
$P_{n,m}$ to oscillate as $m$ varies.  Whether or not these solutions
are relevant hinges on the question of which boundary conditions are
associated with generic physical systems.

\subsection{Remark on boundary conditions}
\label{sec:boundaryconditions}
Given several families of homogeneous solutions, it is natural to ask
which solution will be selected for a given set of boundary
conditions.  From a mathematical perspective, this is a
straightforward question.  As described in detail for the 6-fold case
in Reference~\cite{scs}, the only way to specify boundary conditions
on Eq.~(\ref{eqn:discrete}) that are sure to be self-consistent and
not produce negative chain densities is to specify the densities of
only the ingoing chains at each boundary.  In general, the densities
specified will not match perfectly any of the homogeneous solutions
$P_{n,m}$ for all $m$ and $n$, even if horizontal translational
symmetry is assumed.  Based on experience with the 6-fold case, we
might expect the trajectory in the space of $P_{n,m}$ as the slab is
traversed to quickly approach a fixed point and stay near it over a
substantial portion of the slab, then make a rapid transition to the
vicinity of another fixed point, and finally move off to a point
consistent with the bottom boundary condition.  The details of the
transition regions and selection of the fixed point remain to be
worked out in the present case.

From a physical perspective, the situation is not so clear.  Consider
the following two possible choices of boundary conditions for the top
surface.  In both cases, assume that $P_{n,m}=0$ for all ingoing
$(n,m)$ except the ones specified as follows: (BC1)
$P_{0,5}=P_{1,5}=1$; and (BC2) $P_{0,1}=P_{1,1}=\xi^5$.  Assume also
that the boundary conditions on the bottom slab are simple reflections
of these through the horizontal.  Thus we have boundary conditions
with the same total stress stipulated, but in one case it is injected
via a sparse set of strong forces, and in the other via a denser set
of weaker forces.  These conditions will {\em not} necessarily lead to
the same fixed points in the bulk.  Since the chain densities are
measured with respect to the intrinsic length scale $\lambda$, BC1 and
BC2 are not simply related by a scale transformation.  We therefore
expect them to lead to fixed points corresponding to different $p$'s,
which implies that they will require different rescalings of $f_0$.
This in turn indicates that the force scales fixed by the two
different boundary conditions differ, in spite of the fact that they
correspond to specifying the same overall force.

The resolution of this apparent paradox lies in the fact that the
boundary conditions do not specify the densities of the {\em outgoing}
chains.  Ultimately, the force scale must be determined by the $zz$
component of the stress, which must be constant throughout the slab;
i.e., $\sigma_{zz}$ must determine the prescribed rescaling of $f_0$.
But the densities of outgoing chains produced by BC1 and BC2 will differ,
which will cause $\sigma_{zz}$ to differ in the two cases.  In order
to determine $\sigma_{zz}$, which is the total pressure applied to the
top surface, we must first determine the full solution, or at least
which fixed point the trajectory approaches.  

The converse situation, in which the stress is specified but not the
individual chain densities, is more familiar, but just as subtle.  One
might expect it to be sufficient to specify some components of the
stress tensor and its spatial derivatives on the top boundary and
others on the bottom, but the DFCN equations require independent
specifications of densities of chains of many directions and
strengths.  In an experiment in which a granular material is put under
compression between to flat plates, it is not at all clear how the
chains densities at the boundaries are determined.  Both the
distribution over directions $n$ and strengths $m$ are relevant.  An
important topic of future research will be to develop a method of
assigning directions to force chains observed in experimental images
or numerical models, and to learn from them how the boundary
conditions appear to be imposed.

\section{Conclusion} 
\label{sec:conclusion}
In this paper we have exhibited for the first time solutions of the
\eqname\ equation for the chain densities in a model that permits the
interaction of chains of different intensities.  These solutions have
highly nontrivial structure that raises a number of interesting
questions.  Most importantly, they permit calculations of both
stresses and single particle force distributions in a unified way.
Refinement of the model is clearly necessary: we would like to avoid
divergences in the density of weak chains; and we would like to
generalize the splitting and fusion functions, which might bring the
single contact force distribution ${\mathcal P}(f)$ into closer
agreement with experiment.  Nevertheless, there is much to be gained
from further study of the models presented here, or minor modifications
of them.

The availability of closed form analytic solutions suggests that a
complete theory of the stress configurations and response functions
for these particular models is within reach.  In addition, it appears
possible that the linearized theory in the vicinity of the fixed
points presented here could lead to a new derivation of constitutive
relations for closing the equilibrium equations for the macroscopic
stresses.  Finally, the solutions presented here may provide a basis
for extensions to similar discrete models in which more than one type
of vertex structure is allowed.  All of these topics are currently
under study.\cite{bcoprivate} The author looks forward to continued
collaboration with David Schaeffer on these topics, as he is certain
to offer valuable insights and pose questions that lead to fruitful
calculations.

\section*{Acknowledgments}
I thank J.-P.~Bouchaud, P.~Claudin, M.~Otto, and D.~Schaeffer for
extensive discussions and the group of E.~Cl\'{e}ment and the LMDH at
Jussieu in Paris for its hospitality.  I am especially grateful to
P.~Claudin for his suggestions for improving a draft of this paper.
This work was supported by the NSF Grant PHY-98-70028.


\appendix

\smallskip

Received December 2002.

 \smallskip

\end{document}